\documentclass[prl,twocolumn,amsmath,amssymb,floatfix]{revtex4}

\def\url#1{}

\usepackage{graphicx}
\usepackage{bm}
\usepackage{braket}

\usepackage{bbm}

\usepackage{amsmath}
\usepackage{braket}
\usepackage{graphicx} 
\usepackage{color}

\begin{document}

\title{Incoherent qubit control using the quantum Zeno effect}
\author{S. Hacohen-Gourgy$^{1,2}$}
\author{L. P. Garc\'ia-Pintos$^{3,4}$}
\author{L. S. Martin$^{1,2}$}
\author{J. Dressel$^{3,4}$}
\author{I. Siddiqi$^{1,2}$}
\affiliation{$^1$Quantum Nanoelectronics Laboratory, Department of Physics, University of California, Berkeley CA 94720, USA\\$^2$Center for Quantum Coherent Science, Department of Physics, University of California, Berkeley CA 94720, USA\\$^3$Institute for Quantum Studies, Chapman University, Orange, CA 92866, USA\\$^4$Schmid College of Science and Technology, Chapman University, Orange, CA 92866, USA }

\begin{abstract}
The quantum Zeno effect is the suppression of Hamiltonian evolution by repeated observation, resulting in the pinning of the state to an eigenstate of the measurement observable. Using measurement only, control of the state can be achieved if the observable is slowly varied such that the state tracks the now time-dependent eigenstate. We demonstrate this using a circuit-QED readout technique that couples to a dynamically controllable observable of a qubit. Continuous monitoring of the measurement record allows us to detect an escape from the eigenstate, thus serving as a built-in form of error detection.
We show this by post-selecting on realizations with arbitrarily high fidelity with respect to the target state. Our dynamical measurement operator technique offers a new tool for numerous forms of quantum feedback protocols, including adaptive measurements and rapid state purification.
\end{abstract}

\maketitle
In the field of quantum control, two essentially distinct resources are available for state manipulation. Application of a time-dependent Hamiltonian via external driving enables state preparation given a known initial state. In contrast, measurement and dissipation provide a uniquely quantum resource, owing to the stochastic back-action that necessarily accompanies acquisition of information. In addition, measurement-based, or incoherent control~\cite{wiseman2009quantum} also extracts entropy from a system, this information can be used to detect and correct for errors and imperfections. While incoherent and Hamiltonian control are often used in conjunction~\cite{Felix2015,martin2015deterministic,thomsen2002continuous,JacobsPurification2003,Riste2013,NKatz2008}, full control is also possible using measurement alone~\cite{wiseman2011quantum,Nori2010,Jacobs2010,roa2006measurement,aharonov1980ZenoDrag,Pechen2006,Feng2008}. Measurement-only manipulation has been demonstrated using a fixed measurement basis~\cite{Blok2014}, but unlike Hamiltonian-based methods, implementation of a time-dependent measurement basis is lacking. Such a capability is a versatile additional degree of freedom for measurement based protocols, such as rapid state purification~\cite{JacobsPurification2003} and state manipulation~\cite{DragFB,Nori2010,Jacobs2010}, for control by projection into a subspace referred to as quantum Zeno dynamics~\cite{burgarth2014QZD}, and for measurement-based quantum computation~\cite{MBQComp}.

\begin{figure}[htp!]
{\includegraphics[width=1\linewidth]{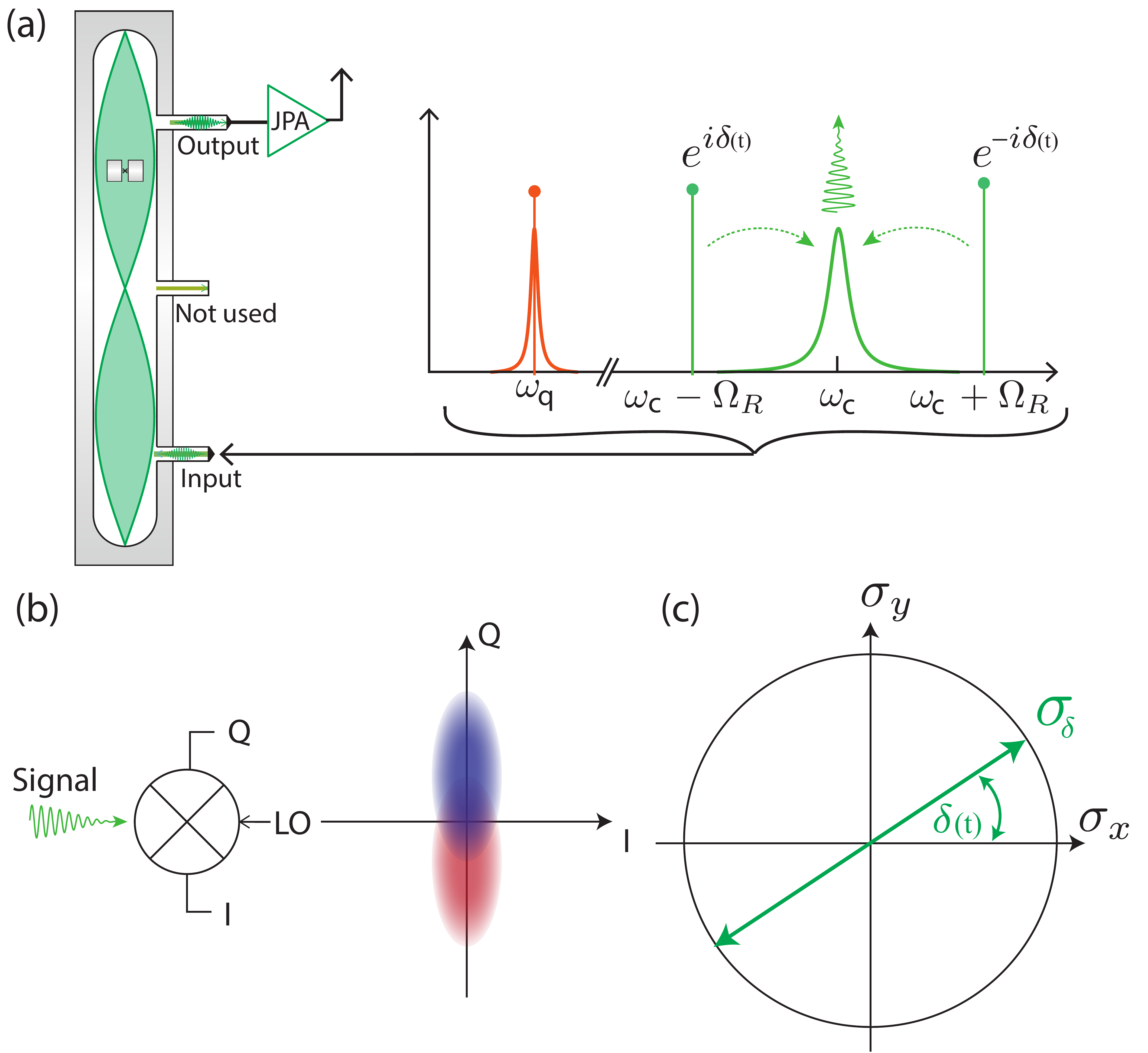}}
\caption{(a)(b) Schematic of the main components in the experimental setup. (a) A transmon qubit in a 3D aluminum cavity. A weakly coupled port is used to simultaneously input three microwave tones; a Rabi drive at $\omega_q$ is used to create an effective low frequency qubit at $\omega_{q-\mathrm{eff}}=\Omega_R=40$MHz (orange), while two sideband tones resonantly couple the effective qubit to the cavity (green). (b) The output signal is amplified with a Josephson Parametric Amplifier (JPA),  and demodulated at room temperature with an IQ mixer. The signal and JPA phase are aligned along the Q quadrature. (c) Illustration of the XY plane in the Bloch sphere of the effective qubit. The green arrow denotes axis of the time dependent measurement operator $\sigma_{\delta(t)}$.}
\label{fig:Setup}
\end{figure}

In this Letter, we present a method to dynamically tune the measurement operator in a circuit-QED system, and use this capability to deterministically and incoherently manipulate the state of an effective qubit. Our method relies on the suppression of coherent evolution via strong measurement, known as the quantum Zeno effect (QZE), which has been observed in many systems~\cite{ZenoPhot1999,ZenoPhotNat2006,ZenoMWPhotHaroche2008,WinelandZeno1990,ZenoAtomic1993,ZenoMol11997,ZenoAtom12001,ZenoAtomRb2006,ZenoSSSpin2013,ZenoSCQubitKorotkov2010,ZenoFluxQ2015,ZenoSlichter}.
Essentially it pins a quantum state to an eigenstate of the measurement operator.
Changing the operator at a rate slow compared to the rate of measurement-induced dephasing $\Gamma_D$, we effectively `drag' the state using measurement alone~\cite{aharonov1980ZenoDrag,Pechen2006,Feng2008,roa2006measurement}. This method does not require the measurement record or feedback to achieve control. However by monitoring the record with a quantum-limited Josephson parametric amplifier (JPA), we characterize the dynamics and verify good agreement with theory. In the fast-driving limit, where the Zeno effect breaks down, we observe a characteristic arcing effect in which the state maintains relatively high purity even as is transitions to the unwanted measurement eigenstate. Using the measurement record to post-select, we show that we can achieve arbitrarily high fidelity with respect to the target state. Thus measurement serves a dual role, both controlling the state and providing real-time information on its performance.\\

Our system setup is similar to the one used in Ref.~\cite{hacohen2016dynamics}. It consists of a transmon~\cite{Koch2007,Transmon3D} qubit dispersively coupled to the modes of a 3D superconducting cavity. We apply a tone resonant with the qubit frequency that drives Rabi oscillations on the qubit at a frequency of $\Omega_R$, so that its Hamiltonian becomes that of an effective qubit with energy splitting determined by $\Omega_R/2 \pi$=40 MHz. 
The new energy eigenstates in this dressed basis are $\ket{\pm}=(\ket{g}\pm\ket{e})/\sqrt{2}$, where $\ket{g}$ and $\ket{e}$ are the ground and excited states of the bare qubit respectively. It is within the frame of this effective qubit that we demonstrate the ability to drag the state. We then apply a pair of sideband tones detuned above and below the cavity frequency by $\Omega_R$, as illustrated in Fig.~\ref{fig:Setup}, which gives us the following Hamiltonian for our effective low frequency qubit~\cite{hacohen2016dynamics},
\begin{equation}
\label{eq:Ham1}
H=\frac{\chi \bar{a}_0}{2} (a+a^\dagger) \sigma_{\delta(t)},
\end{equation}
where $\bar{a}_0$ is the amplitude of sideband tones, $a$, $a^\dagger$ are the cavity ladder operators, and $\chi$ is the qubit dispersive frequency shift.  
The measurement operator $\sigma_{\delta(t)} \equiv \sigma_x \cos \delta(t) +\sigma_y \sin\delta(t)$ is set by the relative sideband phase $\delta(t)$. 
This Hamiltonian is a resonant cavity drive, the sign of which depends on the qubit state along the $\sigma_\delta$ axis. Detecting the cavity output field yields a measurement of the qubit at a rate $\Gamma_M=\Gamma_D \eta = 2 \chi^2 \bar{a}_0^2 \eta/\kappa$ in the $\sigma_\delta$ basis~\cite{Gambetta2008}, where $\kappa$ is the cavity mode decay rate and $\eta$=0.49 is the detection quantum efficiency. We detect the cavity displacement using a JPA operated in phase-sensitive mode, choosing the amplified axis to align with the displaced quadrature. The full system calibration procedure can be found in Ref.~\cite{hacohen2016dynamics}.

\begin{figure}[htp!]
{\includegraphics[width=1\linewidth]{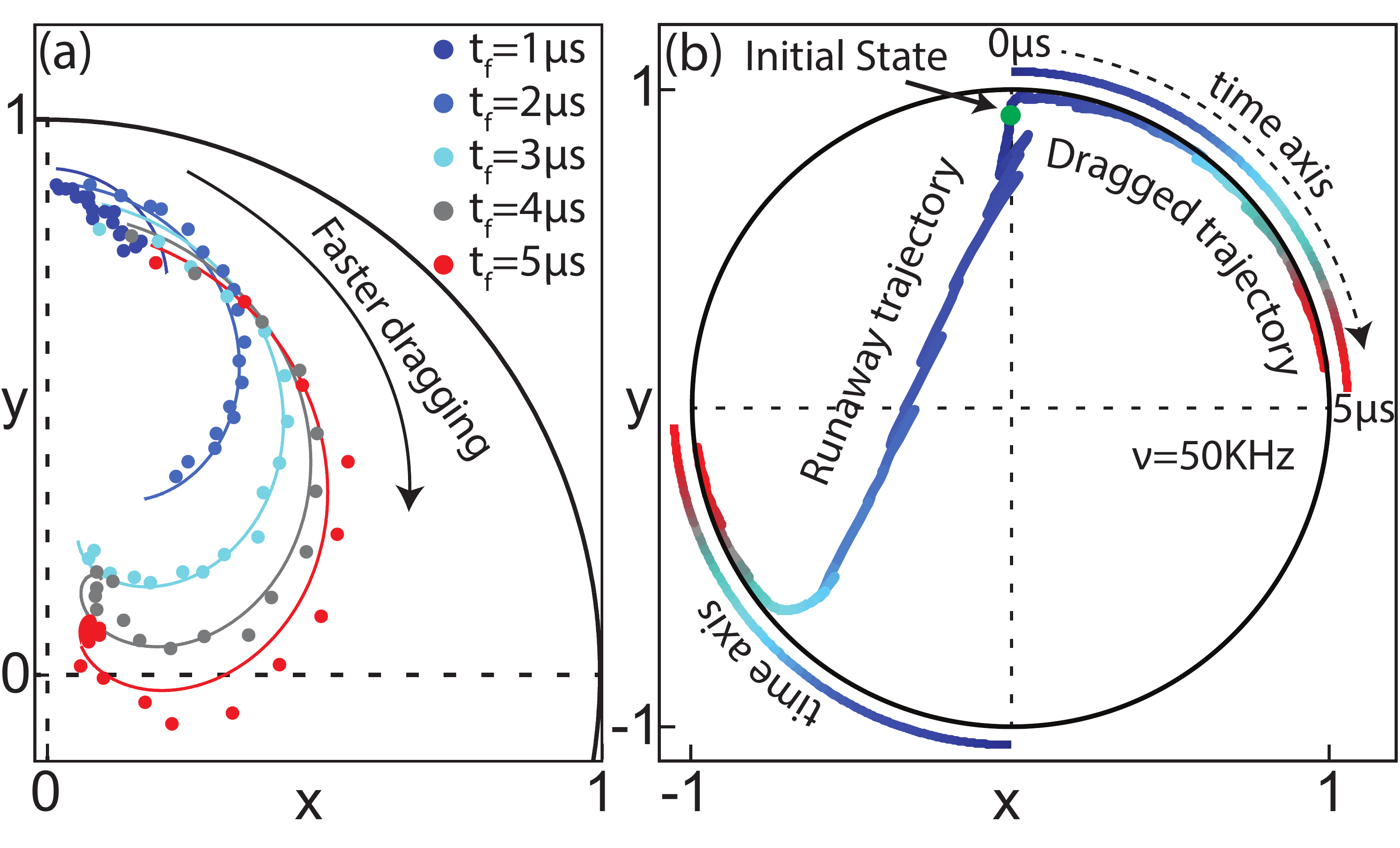}}
\caption{(a) Average state behavior of a qubit being dragged at varying rotation speeds $v$. Dots are tomography results in the XY plane of the Bloch sphere after the fixed time indicated in the legend, and for rotation speeds from 10KHz to 180KHz in steps of 10KHz. Lines are theory plots with the experimental parameters given in the main text. (b) Two example selected trajectories for a dragging rate of $v$=50KHz and a duration of 5$\mu$s: one illustrating successful dragging of the qubit state, whose state remains pure, while the other undergoes a jump and continuous to get dragged along the opposite pole. Colors in the figure correspond to time evolution. The colored lines outside the Bloch sphere indicate the time axis going from blue for t=0$\mu$s to red for t=5$\mu$s, these illustrate the position of the measurement axis as function of time. The same colors correspond to the time evolution of the two trajectories shown. }
\label{fig:AvgBehaviour}
\end{figure}

\begin{figure*}
{\includegraphics[width=1\linewidth]{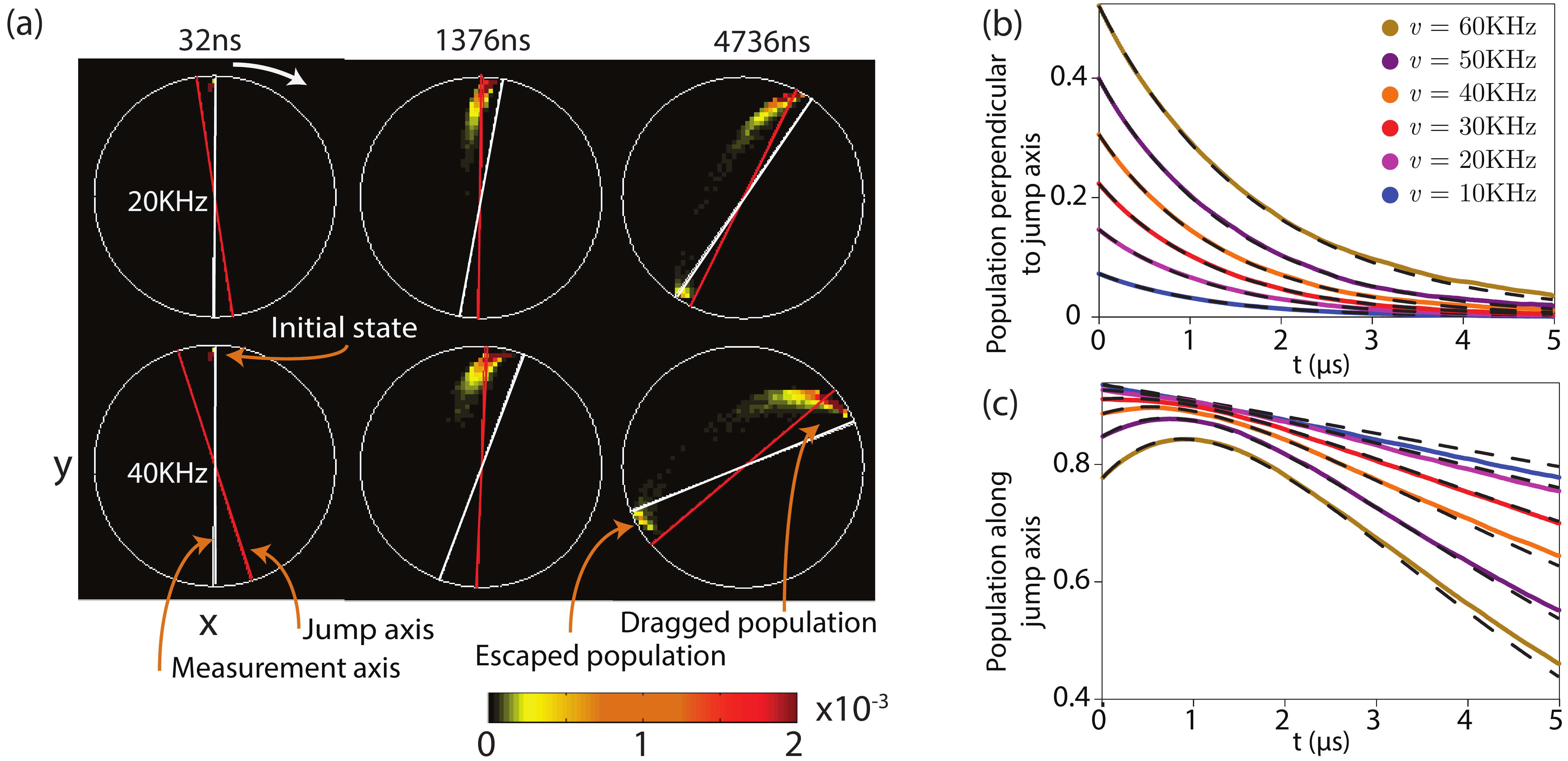}}
\caption{(a) Ensemble histograms of the qubit state as function of time, showing the XY plane of the Bloch sphere, for two example velocities of 20KHz and 40KHz, with $\Gamma_D/2 \pi = 130$KHz. The state is initialized at $|y=+1\rangle$. The measurement axis is represented by white lines, while the theoretically calculated natural jump axis is indicated by red lines. The ensemble average as a function of time in the frame of the jump axis is shown perpendicular to the jump axis (b) and along the jump axis (c). Data are generated by averaging the trajectories as a function of time for the dragging velocities in the Zeno regime $\Gamma_D \ge 2 |\Omega|$. Black dashed lines show theoretical results using the experimental parameters given in the main body.}
\label{fig:DistributionMaps}
\end{figure*}

We start by initializing the effective qubit in the $\ket{y=+1}$ state, which 
corresponds
to the ground state of the non-driven transmon qubit. We then continuously measure the effective qubit while changing the measurement axis. This is followed by one of seven pulses $\{\text{I}, x_{\pi/2}, -x_{\pi/2}, y_{\pi/2}, -y_{\pi/2}, x_{\pi}, -x_{\pi} \}$, and a strong projective measurement for tomography. The dephasing rate during the continuous measurement is fixed, and set to $\Gamma_D/2 \pi$ = 0.13 MHz. We repeat the runs for measurement rotation speeds relative to the effective qubit spanning from $v$=0.01 MHz to $v$=0.18 MHz, and perform tomography at intervals from 1$\mu$s to 5$\mu$s for each rotation speed. 
The thermal population of the transmon qubit was about 15\%, so before each measurement we perform a 1$\mu$s projective measurement heralding 
the preparation state.
We also use the projective readout at the end to ensure that the transmon qubit is still within the two-level subspace after the run. The tomography for the ensemble average behavior is shown in Fig.~\ref{fig:AvgBehaviour}a. The colored dots show the tomography from $\sim$20000 traces per dot, and the lines are theory for the following parameters: initial state with $\langle y \rangle $=0.94, $\langle x \rangle = \langle z \rangle $=0, $\Gamma_D/2 \pi$ = 0.13 MHz, 
and an additional pure dephasing, which we attribute mainly to instabilities in the Rabi drive, at a rate $\Gamma_\phi/2 \pi$=0.005 MHz (corresponding to the decay time of the Rabi oscillations of the bare qubit). 
The statistical errors are negligible and the small discrepancy of the tomography data with theory is most likely due to systematic drifts of the measurement rate (amplitude of the side band tones) and leakage tone at the cavity mode frequency (LO leakage - see methods in Ref~\cite{hacohen2016dynamics}).

We now focus on the conditional dynamics of the state as it is being dragged. For this, we reconstruct the quantum trajectories~\cite{Murch2013,hacohen2016dynamics} from the continuous traces (see also the supplemental material). 
Our system is in a regime where $\kappa \gg \Gamma_D$, in which we can infer the diffusive nature of the quantum jumps. Because we operate the JPA such that it amplifies the optimal (informational) quadrature, the qubit evolution due to the measurement is not affected by phase back-action~\cite{Gambetta2008,Korotkov2001Backaction}.
Then the dynamics of the system can be described by the following master equation, in It\^o form~\cite{wiseman2009quantum,jacobs2006straightforward}:
\begin{equation}
\label{eq:ME}
d\rho = \frac{\Gamma_D}{2} \mathcal{L}[\sigma_{\delta(t)}] \rho ~dt + \sqrt{\frac{\Gamma_D}{2} \eta}~ \mathcal{H}[\sigma_{\delta(t)}] \rho ~dW,
\end{equation}
where $\mathcal{L}[X]\rho = X\rho X^\dagger - (X^\dagger X\rho + \rho X^\dagger X)/2$ is the Lindblad dissipation superoperator, $\mathcal{H}[X]\rho = X\rho + \rho X^\dagger-\langle X\rho + \rho X^\dagger\rangle \rho$, and $dW$ is a Gaussian distributed variable with a variance $dt$~\cite{Oksendal2003}, which is itself extracted from the measurement record. 
We use the POVM that generates this equation with additional corrections to account for extra dephasing on the effective qubit (at a rate $\Gamma_\phi$) to reconstruct the trajectories as function of time from the continuous traces (see supplemental material). Fig.~\ref{fig:AvgBehaviour}b shows two example trajectories for a dragging velocity of $v$=50KHz, with one trajectory showing a state that was successfully dragged, while the other illustrates a `quantum jump'. 
Note that after the jump the measurement process continues to drag the state on the opposite side of the Bloch sphere.

The dynamics of the whole ensemble can be visualized by plotting the 
distribution of the state of the qubit in the Bloch sphere
as function of time, as shown in Fig.~\ref{fig:DistributionMaps}. 
There are several prominent qualitative features in these plots. 
As expected, the rate at which the qubit jumps is larger for faster dragging velocities. 
Strikingly, these quantum jumps always diffuse in an arc that extends 
opposite to
the direction of rotation. This can be understood from the form of the back-action,
which is zero at the poles of the measurement axis,
and maximal in-between. Hence, when the state gets `pushed forward' (that is, in the direction of the rotation) by the back-action, it is pushed towards a region of lower back-action. At the same time, it cannot go past the measurement axis because the back-action goes to zero at the pole. On the other hand, if the state gets `pulled back' by the back-action, it is towards a region of higher back-action, thus having an increased probability of `escaping' and undergoing a transition  to the other side of the Bloch sphere, i.e. a quantum jump. 
Due to the relatively high quantum efficiency of our system, the state remains close to the surface of the Bloch sphere, and trajectories that jump arc out before arriving at the other side. 

A consequence of the arcing feature in the dynamics is the lagging of the average of the state behind the measurement axis. For our specific experiment the ensemble averaged dynamics can be solved analytically by going into a frame rotating at the dragging velocity $v$, where the measurement axis is fixed and the qubit is driven by the Hamiltonian $H = (\Omega / 2) \sigma_z$,
with $\Omega = 2 \pi v$. 
In this measurement-axis frame the average qubit state evolves
according to
\begin{align}
\label{eq:masterIto}
d\rho &= -i \frac{\Omega}{2}[\sigma_z,\rho] ~dt + \frac{\Gamma_D}{2} \mathcal{L}[\sigma_y] \rho ~dt,
\end{align}
where the measurement axis is now fixed along the $y$ direction, and for simplicity we drop the negligible purely dephasing term $\Gamma_\phi$.
The solutions display two characteristically different regimes:
\begin{itemize}
\item[(i)] $\Gamma_D < 2 |\Omega|$ --  oscillatory with $\lambda_\pm$ complex, and
\item[(ii)] $\Gamma_D \ge 2 |\Omega|$ -- overdamped with $\lambda_\pm$ real.
\end{itemize}
$\lambda_\pm = (-\Gamma_D  \pm  \sqrt{ \Gamma_D^2 - 4 \Omega^2})/2$ and $\vec{V}_\pm = \left( 1 , (\lambda_\pm + \Gamma_D)/\Omega \right)$ are the eigenvalues and eigenvectors respectively. In the oscillatory regime the state of the qubit oscillates with respect to the measurement axis, and thus is not dragged by the measurement.
In the overdamped regime, or Zeno regime,
the oscillatory behaviour vanishes and is replaced by exponential decay along the axes defined by the eigenvectors $V_\pm$. 
As $\Gamma_D \rightarrow \infty$, the eigenvalue $\lambda_+$ goes to zero, which means that if the qubit starts near a pole of $\vec{V}_+$ it will remain pinned to it for an arbitrarily long time.
\begin{figure}[t]
{\includegraphics[width=1\linewidth]{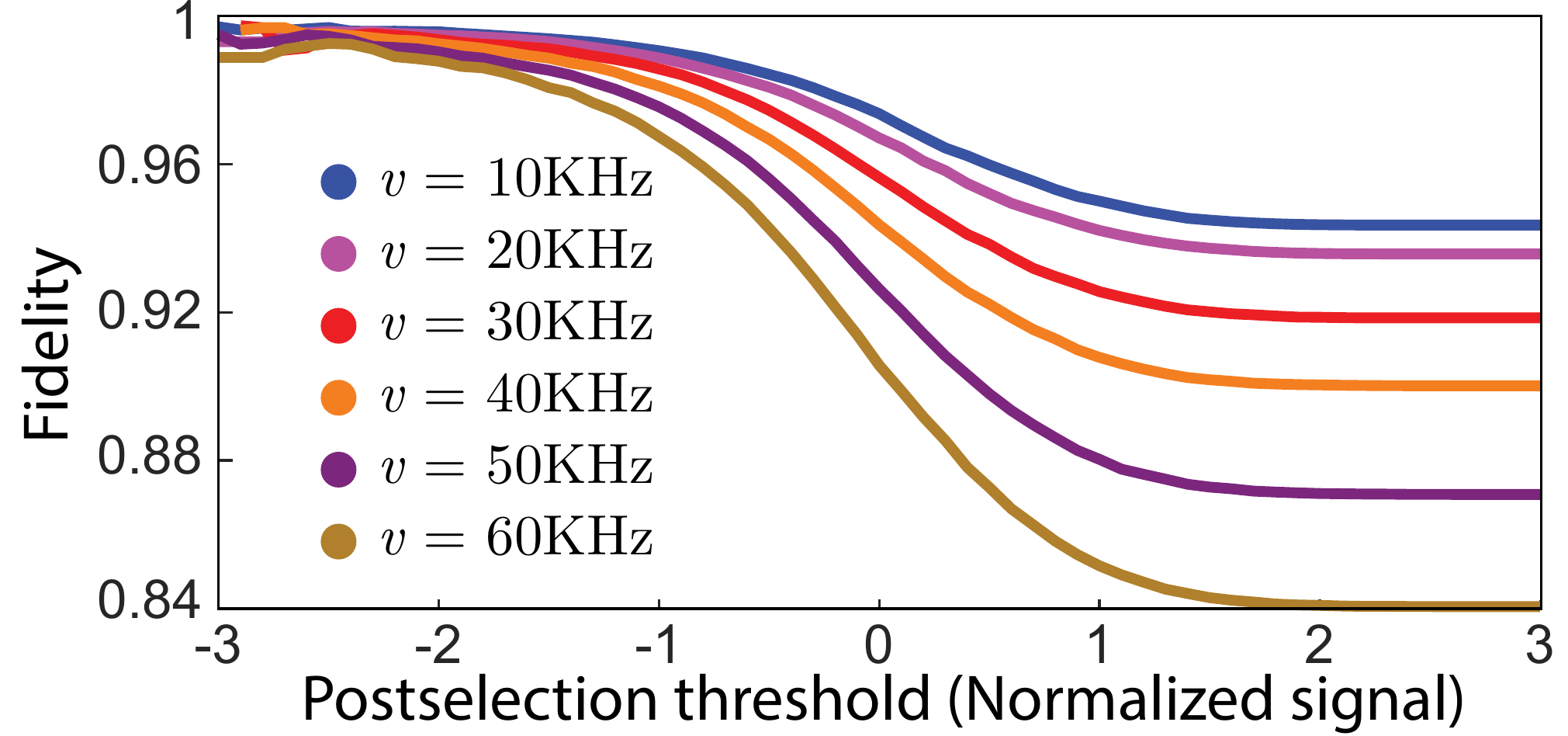}}
\caption{Fidelity as function of post-selection threshold. Post-selection is performed with respect to the average voltage of the detector signal. Each trace corresponds to a fixed dragging rate and shows fidelity with respect to the eigenstate of the measurement operator after rotating for 4$\mu$s. X-axis is the normalized post-selection threshold, normalized such that $\pm 1$ correspond to the average values of the signal for the $\pm 1$ eigenstates of the measurement operator.}
\label{fig:Fidelity}
\end{figure}
The slow decay for $\Gamma_D < \infty$ can be attributed to quantum jumps between the poles of $\vec{V}_+$. In each realization of the experiment these jumps can be observed. The jump axis is identified with the direction in which the damping rate is smallest, since the fast damping in the orthogonal direction aligns the poles of the jump with the slow axis.

Since $\lambda_+ \ge \lambda_-$, the jump axis is the eigenvector $\vec{V}_+$, with a characteristic angle relative to the measurement axis
\begin{align}
\theta =  \arctan\left( \frac{2 \Omega}{\Gamma_D  +  \sqrt{ \Gamma_D^2 - 4 \Omega^2}}   \right).
\end{align}
This angle characterizes the direction along which the population of the qubit concentrates, and is only defined within the Zeno regime, where dragging occurs. 
In such regime a qubit state close to a pole of the jump axis eventually jumps to the other pole at a rate $\gamma_J =|\lambda_{+}|/2$. 
Note that for slow dragging velocities, in the limit $\Gamma_D \gg 2|\Omega|$, the jump axis aligns with the measurement axis, and the jump rate converges to the familiar form $\Omega^2/(2\Gamma_D)$~\cite{Gambetta2008,ZenoAtomic1993,CookZenoJump1988}.

Fig.~\ref{fig:DistributionMaps}a illustrates the jump axis, indicated by a red line, lagging  behind the measurement axis at an angle $\theta$.
Moreover, Fig.~\ref{fig:DistributionMaps}b and Fig.~\ref{fig:DistributionMaps}c show good agreement between theoretical and experimental ensemble dynamics in the frame of the jump axis for dragging velocities in the Zeno regime. 
We can see the exponentially decaying behaviour in the direction perpendicular to the jump axis, indicating that the population is aligning with it. 
The lagging angle between the average state and the measurement axis can be understood to arise from competition between the stochastic back-action and rotation. 

Without observing the measurement outcome there is an optimal initial measurement axis and rotation velocity that maximizes the fidelity with respect to a target state~\cite{Feng2008}.
However, as the magnitude of the back-action depends on the measurement outcome, its relative size can be inferred from the measurement record. As a larger positive measurement outcome induces a larger change toward the measurement axis, one can use this effect to post-select on trajectories in which the state was pulled closer to the measurement axis. In Fig.~\ref{fig:Fidelity}, we show fidelity with respect to the target measurement eigenstate for various post-selection criteria. One can see that the more aggressively one post-selects on the integrated voltage, the higher the resulting fidelity. The apparent degradation in fidelity for the most negative postselections is due to insufficient statistics. Thus, measurement allows us not only to drag the state, but also to monitor its dynamics and herald arbitrarily high fidelity. The above dynamics suggest that given a `runaway' state, or an `error', the measurement axis could rotate and drive it back via a feedback protocol, achieving improved control. A feedback protocol achieving such a result has been shown~\cite{DragFB}. The idea is to feedback on the measurement axis such that it is always half way between the current state and the target state.




This dynamical control of the measurement operator enables novel capabilities for qubit control, such as the incoherent control demonstrated here, improved incoherent control with feedback~\cite{DragFB}, rapid state purification~\cite{JacobsPurification2003,wiseman2006reconsideringPurification,combes2010rapid}, and adaptive measurements~\cite{wiseman2009adaptive,wiseman2009quantum}. This measurement scheme also generalizes to multi-level systems. 
In such multi-level settings, fast measurement rates of certain operators restrict the system to evolve within a particular subspace of the total Hilbert space, which is known as Quantum Zeno Dynamics~\cite{QZD1,QZDExp1,QZD2}. Such restriction has been recently shown to enable universal quantum computation within that subspace~\cite{Burgarth2014}. Changing these subspaces dynamically through the evolution of the monitored operators is an avenue that has yet to be explored.

\textit{Acknowledgments} We thank Alexander Korotkov, Andrew Jordan, Juan Atalaya, Emanuel Flurin, Machiel Blok and Howard Wiseman for discussions. L.S.M. acknowledges support from the National Science Foundation (graduate fellowship grant 1106400). and from a Berkeley Fellowship for Graduate Study. This work was supported by the Air Force Office of Scientific Research (grant FA9550-12-1-0378) and the Army Research Office (grant W911NF-15-1-0496).






\bibliography{RSbib}
\section{Supplemental Material}

\subsection{Device parameters}
The qubit has a transition frequency of $\omega_q/2\pi$ = 4.262 GHz, an energy-decay timescale of $T_1 = 60~\mu\mathrm{s}$, and a dephasing (Ramsey decay) timescale of $T_{2}^{*} = 30~\mu\mathrm{s}$. For this experiment we use the second lowest cavity mode, with a frequency of $\omega/2 \pi$ = 7.391 GHz, and a linewidth of $\kappa/2 \pi$ = 4.3 MHz. The qubit dispersive frequency shift is $\chi/2 \pi$ = -0.23 MHz.

\subsection{Trajectory reconstruction}
The stochastic master equation given in the main text is generated by the following measurement operator
\begin{align}
\label{eq:MeasOpp}
\Omega(V) &= \exp \left[-\frac{\Gamma_{D} \eta}{2} \left(V(t)-\sigma_{\delta(t)}\right)^2 dt \right] \\ \nonumber
\rho(t+dt) &= \mathcal{E}_{1-\eta}\frac{\Omega \rho(t) \Omega^\dagger}{\text{Tr}[\Omega \rho(t) \Omega^\dagger]}, \\ \nonumber
\end{align}
where $\mathcal{E}_{1-\eta_i}$ is a superoperator which models dephasing due to finite quantum efficiency and small additional dephasing taken from the finite measured Rabi time. To ensure positivity of the state when $dt$ is taken to be finite, we use Eq.~(\ref{eq:MeasOpp}) to numerically propagate the quantum trajectories. The parameters $\Gamma_D$ and $\eta$ are calibrated independently. The former we measure by preparing $|+\rangle$ and then performing a Ramsey measurement. We measure the quantum efficiency by preparing states $|y=\pm 1\rangle$. Histograms of the integrated measurement records yield a pair of Gaussians which separate as a function of time. The quantum efficiency is given by
\begin{equation}
\eta = \frac{(\mu_{y=+1} - \mu_{y=-1})^2}{8 \tau \sigma^2 \Gamma_{D} },
\end{equation}
where $\mu_{y=\pm 1}$ is the mean of the Gaussian for the $|y=\pm 1\rangle$ state preparation, $\sigma$ is the average standard deviation of the Gaussians and $\tau$ is the measurement duration~\cite{Korotkov2014,Korotkov2016}.

\begin{figure}[htp!]
{\includegraphics[width=1\linewidth]{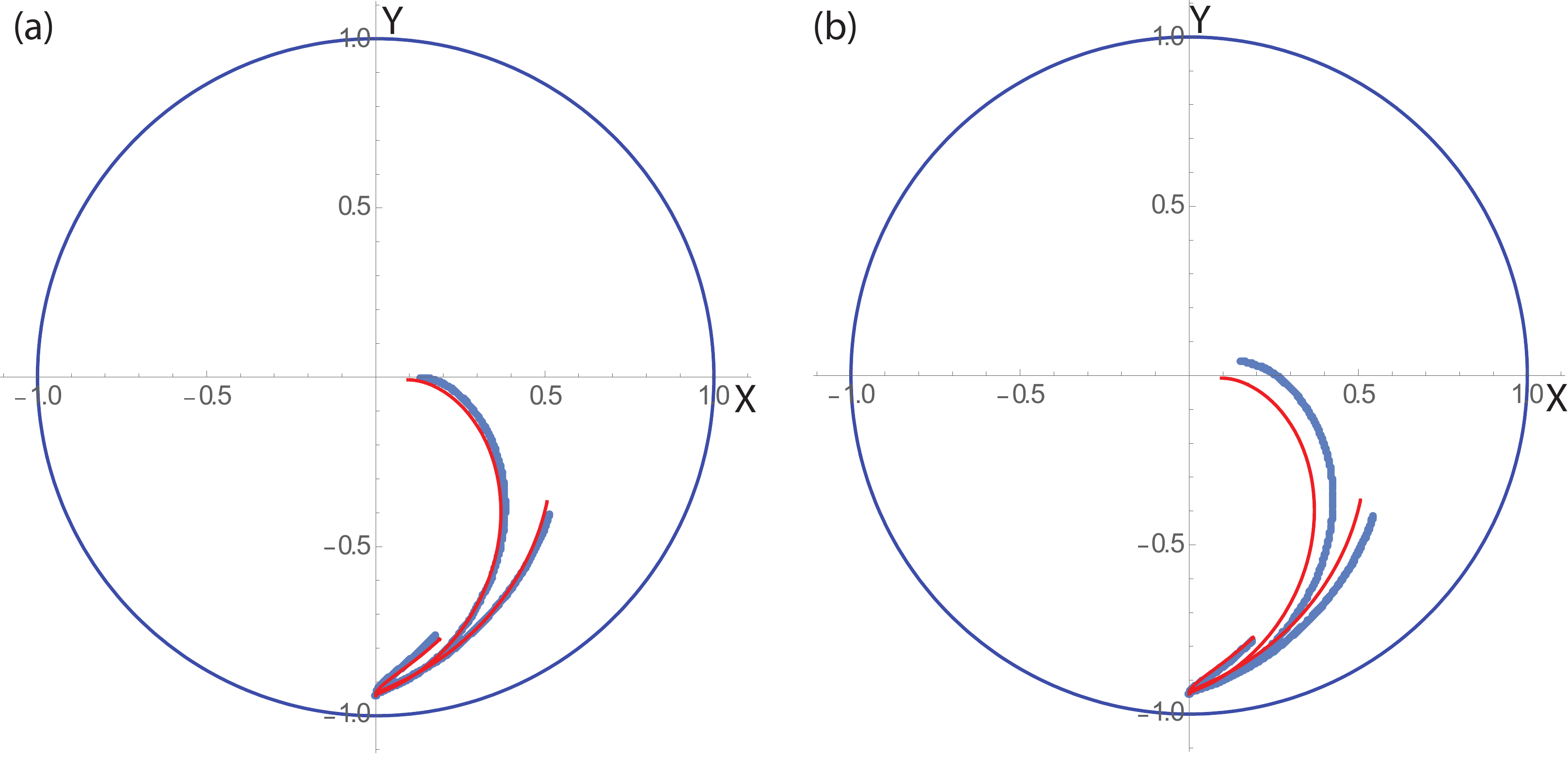}}
\caption{Average of the experimental trajectories before (b) and after (a) correction of offset in detector voltage, compared to theory for 3 example dragging velocities of 10KHz, 40KHz and 100KHz. Blue is the average of the experimental trajectories and the red line is the solution to the master equation. }
\label{fig:Offset}
\end{figure}

When reconstructing the quantum trajectories and comparing the average of the trajectories to the solution for the master equation for the the average state we found a disagreement between theory and experiment. This seemed to be a systematic discrepancy due to a small offset in the detector voltage. We corrected this using an informed `guess' offset, which was calibrated from a different experiment~\cite{Atalaya2017} performed using the same setup. The offset value used is 0.17V where the separation of the mean of the Gaussians for this detector was 1.74V. In Fig.~\ref{fig:Offset} we show the comparison of the average of the trajectories with theory for 3 dragging velocities, for processing with and without the correction. In the main text we use the corrected data.
\begin{center}
\begin{figure*}
\includegraphics[width=0.8\linewidth]{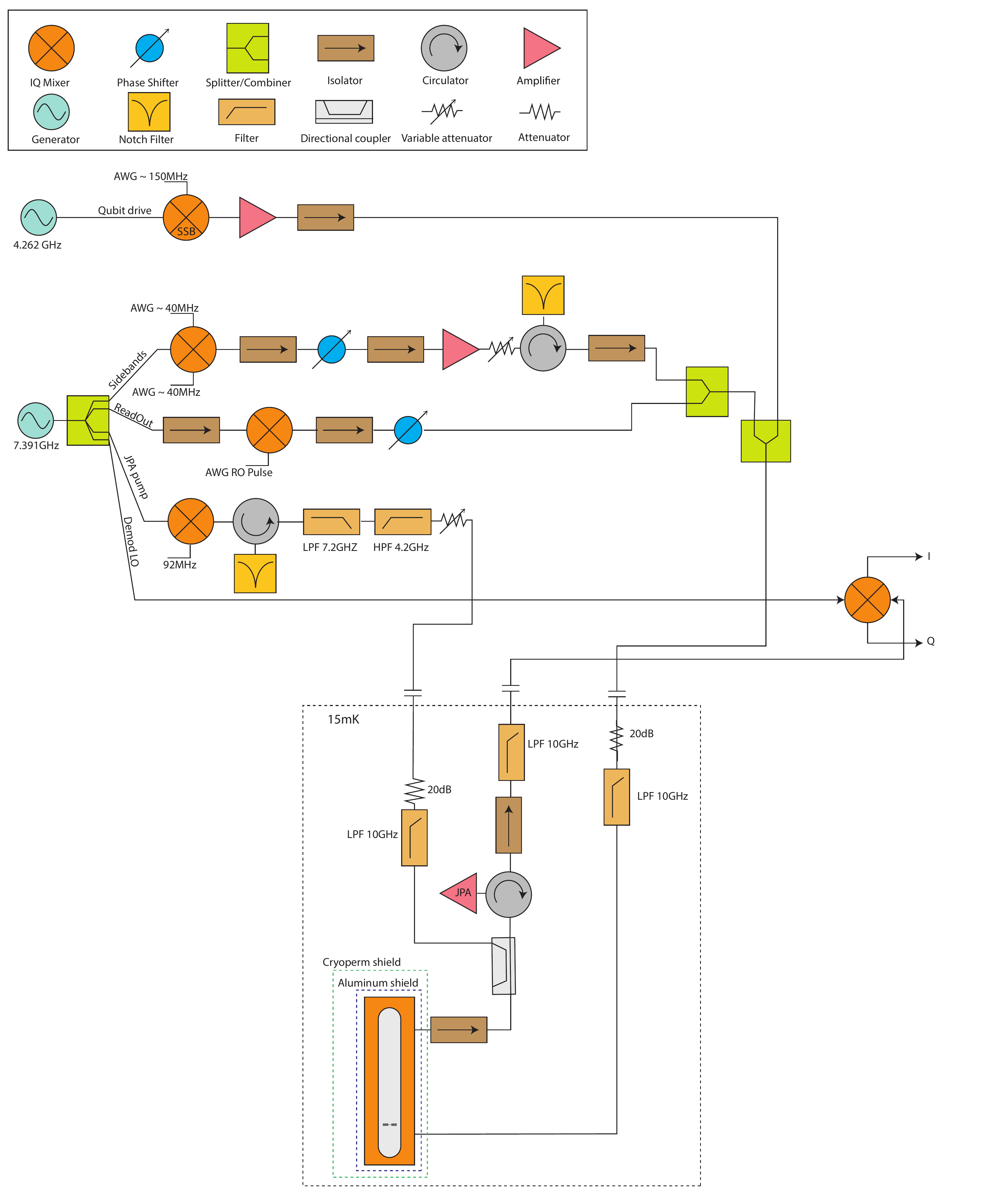}
\label{fig:SetupDiag}
\caption{Schematic illustration of the experimental setup}
\end{figure*}
\end{center}
\end{document}